\documentclass[fleqn,usenatbib]{mnras}

\usepackage{newtxtext,newtxmath}

\usepackage[T1]{fontenc}

\DeclareRobustCommand{\VAN}[3]{#2}
\let\VANthebibliography\thebibliography
\def\thebibliography{\DeclareRobustCommand{\VAN}[3]{##3}\VANthebibliography}

\usepackage{graphicx}	
\usepackage{amsmath}	
\usepackage{amssymb}	
\usepackage{lineno}

\newcommand{\hess}{HESS\,J1023-575}
\newcommand{\psr}{PSR\,J1023--5746}
\newcommand{\ext}{FGES\,J1023.3--5747}

\title[Westerlund 2 in gamma-rays]{Probing the hadronic nature of the gamma-ray emission associated with Westerlund 2}
    
\author[Enrique Mestre et al.]{Enrique Mestre$^{1,2}$,
Emma de O\~na Wilhelmi$^{1,2,3}$\thanks{E-mail: Emma.de.ona.wilhelmi@desy.de}, Diego F. Torres$^{4,1,2}$, Tim Lukas Holch$^{5}$,
\newauthor
Ullrich Schwanke$^{5}$,
Felix~Aharonian$^{6,7,8}$, Pablo Saz Parkinson$^{9,10}$, Ruizhi Yang$^{11,12,13}$,
\newauthor
Roberta Zanin$^{14}$
\\
$^{1}$Institute of Space Sciences (ICE/CSIC), Campus UAB, Carrer de Can Magrans s/n, 08193 Barcelona,
Spain\\
$^{2}$Institut d'Estudis Espacials de Catalunya (IEEC), 08034 Barcelona, Spain \\ 
$^{3}$Deutsches Elektronen Synchrotron DESY, 15738 Zeuthen, Germany\\
$^{4}$Instituci\'o Catalana de Recerca i Estudis Avan\c cats (ICREA), E-08010
Barcelona, Spain \\
$^{5}$Humboldt University of Berlin, Newtonstr. 15, 12489 Berlin, Germany\\
$^{6}$Dublin Institute for Advanced Studies, 31 Fitzwilliam Place, Dublin 2, Ireland \\
$^{7}$Max-Planck-Institut f\"ur Kernphysik, P.O. Box 103980, D 69029 Heidelberg, Germany \\
$^{8}$Gran Sasso Science Institute, 7 viale Francesco Crispi, 67100 L'Aquila,  Italy\\
$^{9}$Department of Physics and Laboratory for Space Research, The University of Hong Kong, Pokfulam Road, Hong Kong\\
$^{10}$Santa Cruz Institute for Particle Physics, University of California, Santa Cruz, CA, 95064, USA\\
$^{11}$Department of Astronomy, School of Physical Sciences, University of Science and Technology of China, Hefei, Anhui 230026, China\\
$^{12}$ CAS Key Labrotory for Research in Galaxies and Cosmology, University of Science and Technology of China, Hefei, Anhui
230026, China\\
$^{13}$School of Astronomy and Space Science, University of Science and Technology of China, Hefei, Anhui 230026, China\\
$^{14}$CTA Observatory GmbH, Via Piero Gobetti 93, I-40129 Bologna, Italy
}

\date{Accepted 2021 May 18. Received 2021 May 18; in original form 2021 February 10}

\pubyear{2015}

\begin{document}
\label{firstpage}
\pagerange{\pageref{firstpage}--\pageref{lastpage}}
\maketitle

\begin{abstract}
Star-forming regions have been proposed as potential Galactic cosmic ray accelerators for decades. Cosmic ray acceleration can be probed through observations of gamma-rays produced in inelastic proton-proton collisions, at GeV and TeV energies. In this paper, we analyze more than 11 years of \emph{Fermi}-LAT data from the direction of Westerlund 2, one of the most massive and best-studied star-forming regions in our Galaxy. In particular, we investigate the characteristics of the bright pulsar \psr\, that dominates the gamma-ray emission below a few GeV at the position of Westerlund 2, and the underlying extended source \ext. The analysis results in a clear identification of \ext\ as the GeV counterpart of the TeV source \hess, through its morphological and spectral properties. This identification provides new clues about the origin of the \hess\ gamma-ray emission, favouring a hadronic origin of the emission, powered by Westerlund 2, rather than a leptonic origin related to either the pulsar wind nebula associated with \psr\ or the cluster itself. This result indirectly supports the hypothesis that star-forming regions can contribute to the cosmic-ray sea observed in our Galaxy.
\end{abstract}




\section{Introduction}

The potential of massive star clusters to accelerate Galactic cosmic rays (GCRs) to very-high energies (VHE, E > 100 GeV) has been recognized since the 1980s \citep{1983SSRv...36..173C,2020SSRv..216...42B,2005ApJ...634..351B}. Several hypotheses have been proposed for acceleration sites to very high energies in star-forming regions (SFRs), either in the vicinity of OB and WR stars, or at the interaction of their fast winds with supernova (SN) shocks, or in so-called super-bubbles (see e.g. \citealt{2020SSRv..216...42B,2019IJMPD..2830022G} and references therein). The presence of cosmic rays (CRs) can be inferred by means of gamma-ray observations (above a few hundreds of MeV), by looking at the by-product of inelastic proton-proton collisions. The spectral energy distribution (SED) from hadronic-originated gamma-ray sources is characterized by a sharp rise in the $\sim$70--200 MeV range (resulting from the neutral pion threshold production energy), followed by a hard emission up to the maximum energy. Competing gamma-ray processes related to leptonic emission should have different imprints on the spectral shape, showing a (in many occasions broad) peak in the hundreds of GeV range. Therefore, sampling the spectrum from a few hundreds of MeV to tens of TeV should result in a strong indication of the origin of the observed radiation. 
Several SFRs have been identified as likely GCR accelerators in the GeV and TeV regime \citep{2011Sci...334.1103A,2011A&A...525A..46H,2012A&A...537A.114A,2015Sci...347..406H,2018A&A...611A..77Y,2019NatAs...3..561A,2020Saha,2020A&A...640A..60Y,2020MNRAS.494.3405S}. However, a firm identification remains elusive, given the large extension of the sources and/or the presence of some other efficient accelerators in the region, such as pulsars and pulsar wind nebulae (PWNe). 

One of the most massive and well-studied SFRs in our Galaxy is Westerlund\,2. The cluster itself presents a $\sim$0.2$^{\rm o}$ wind-blown bubble observed in infrared (IR) by {\it Spitzer} and in radio continuum by ATCA \citep{1997A&A...317..563W,2004ApJS..154..315W}. This bubble is coincident with a prominent feature at radio wavelengths, known as the \emph{blister}, and with an extended TeV source ($\sim$0.2$^{\rm o}$) known as \hess\ \citep{2007A&A...467.1075A,2011A&A...525A..46H}. In the GeV regime, the young and energetic pulsar \psr\ lies 8\,arcmin away from the cluster, and its PWN was initially suggested as possible counterpart of the TeV source \citep{2010ApJ...725..571S,2011ApJ...726...35A} (see Table \ref{tab:pwn} for the physical parameters of the pulsar, obtained from \citealt{2010ApJ...725..571S}). Besides the PWN scenario, several accelerators have been proposed to power the TeV emission: Westerlund 2 contains an extraordinary ensemble of hot and massive OB stars \citep{2004A&A...420L...9R}, for which a total energy release for the collective winds was estimated to be $\sim5.7\times10^{37}$ erg s$^{-1}$. Westerlund\,2 also hosts one of the most massive binary systems, composed of two WN6ha stars \citep{2004A&A...420L...9R}, dubbed WR~20a. Their orbital period is $\sim3.6$ days \citep{2004ApJ...611L..33B,2004A&A...420L...9R} and the estimated kinetic energy loss rate is $\sim10^{37}$ erg s$^{-1}$. Several massive molecular clouds were found within the surrounding of Westerlund 2, overlapping with the GeV and TeV gamma-ray sources (see Fig. 5 in \citealt{2011A&A...525A..46H}).

Since the discovery of the Westerlund 2 cluster, there have been several attempts to determine accurately its distance, estimated between 2 kpc and 8 kpc \citep{1955ApJ...121..161S,1991AJ....102..642M,1997yCat..41270423P,2004MNRAS.347..625C, 2005ApJ...629..512U,2005A&A...432..985R, 2007ApJ...665L.163D,2007ApJ...665..719T,2007A&A...463..981R,2007A&A...466..137A,2009ApJ...696L.115F,2013AJ....145..125V}. In the following we adopt a value of 5 kpc, roughly in the mid-range of the recent optical photometric work \citep{2013AJ....145..125V, 2018MNRAS.480.2109D} and at the bottom end of the interstellar medium (ISM) estimates \citep{2007ApJ...665L.163D,2014ApJ...781...70F}.

The Large Area Telescope (LAT, \citealt{2009ApJ...697.1071A}) onboard the \emph{Fermi Gamma-ray Space Telescope}, launched in June 2008, surveys the gamma-ray sky in the 20 MeV to greater than 300 GeV energy range. The LAT data recorded during the last decade has resulted in different catalogues of gamma-ray sources \citep{2020Ballet}. For instance, the 4FGL catalogue \citep{Abdollahi_2020} covers the data recorded during the first 8 years of observations. On the other hand, the extended sources are listed in the \emph{Fermi} Galactic Extended Source (FGES) Catalog \citep{2017ApJ...843..139A}. In this paper, we analyzed the gamma-ray emission towards \ext\ \citep{2017Ackermann} and the pulsar \psr\ taking advantage of more than 11 years of \emph{Fermi}-LAT data and the most recent gamma-ray source catalogue released by the \emph{Fermi}-LAT collaboration. The paper is structured as follows. In Sections \ref{sec2} and \ref{sec3} we describe the collected data and present the results of the analysis. In Section \ref{sec4} we discuss the possible leptonic (\ref{sec4.1}) or hadronic (\ref{sec4.2}) origin of the \ext\ gamma-ray extended emission. Finally, in Section \ref{sec5} we summarize the conclusions reached.


\section{Data analysis}
\label{sec2}

To investigate the characteristics of the gamma-ray emission from \ext\ and \psr, we used \emph{Fermi}-LAT (P8R3, \citealt{Atwood:2013rka}; \citealt{bruel2018fermilat}) data spanning from 2008 August 4 to 2019 April 24 (or 239557417 - 577782027 seconds in \emph{Fermi} Mission Elapsed Time), in the energy range from 200\,MeV to 500\,GeV. We retrieved the data from a region of interest (ROI) defined by a radius of 20\degr\ around the position of \psr\ \citep[RA= $155.76\degr$, DEC = $-57.77\degr$, ][]{2015ApJ...814..128K}. 
We analyzed only the dubbed SOURCE class events with a maximum zenith angle of 90\degr{} to eliminate Earth limb events.
The events were selected with a minimum energy of 200\,MeV, to avoid events poorly reconstructed due to the large angular resolution and the large crowding of sources in the region. 
The analysis of the source spectrum below 200\,MeV is crucial to characterize hadronic-originated gamma-rays sources, since neutral pion decay spectrum rises steeply below this energy, a feature often referred to as the pion-decay bump. However, this particular source is not a promising candidate for this analysis: the moderate flux of the source and hard spectrum limits the number of photons in the very low energy regime; the extended morphology complicates the analysis with respect to the point-like case; the presence of several sources in the region may contaminate further the spectrum especially at low energies, preventing an accurate characterization (i.e., overlapping with the \psr\ emission). In this work, we focus on studying the general GeV to TeV spectral shape.

To obtain the spectrum and morphology of \ext, we first had to disentangle its diffuse gamma-ray emission from the radiation coming from the pulsar \psr\ and other nearby sources.  For that purpose, we need to derive first the spectral energy distribution of all the sources in the field of view. Therefore, a comprehensive model describing the gamma-ray sources in the ROI is needed. In order to build this model, we included all the LAT sources listed in the \emph{Fermi}-LAT Fourth Source Catalog (4FGL, \citealt{Abdollahi_2020}) in a radius of 20\degr{} around the position of \ext. On the other hand, the LAT data contain a significant contribution from Galactic and extra-galactic diffuse gamma-rays, which is described with the latest version available of the Galactic (gll\_iem\_v07) and isotropic (iso\_P8R3\_SOURCE\_V2\_v1) diffuse emission models. The model's free parameters correspond to the ones of the sources within 3\degr{} around the position of \psr. Beyond this radius, the normalization of all the sources with test statistic greater than ten ($\rm{TS} > 10$) is also free. The Test Statistic is defined as $\rm{TS} = 2log(L/L_{0})$, where $L$ is the maximum value of the likelihood function over the ROI including the source in the model and $L_{0}$ is the same without the source \citep{1996ApJ...461..396M}. Hence, the detection significance of a source is usually approximated by the square root of the TS.

The analysis of the LAT data described above was performed by means of the \textsc{fermipy} \textsc{python} package (version 0.18.0), based on the \textsc{Fermi Science Tools} \citep{2017ICRC...35..824W}. The data were binned in 8 energy bins per decade and spatial bins of 0.1\degr{} size to perform the analysis. The response of the LAT instrument is evaluated with the Instrument Response Functions (IRF, version P8R3\_SOURCE\_V2). The energy dispersion correction was applied to the sources in our model, except for the isotropic diffuse emission model.  

The current version of LAT data (i.e PASS 8) classifies the events into quartiles based on the varying quality of the direction reconstruction (PSF events types). In our analysis, we took into account the four Point Spread Function (PSF) event types available (dubbed PSF0, PSF1, PSF2, and PSF3), and applied the appropriate IRF and isotropic background models according to the quality of the reconstructed event directions. Doing so, we are preventing the loss of possibly useful information in the analysis by means of the separate treatment of high-quality events and poorly localized ones. The analysis is performed by means of a joint likelihood fitting process.

In order to discriminate the diffuse gamma-ray emission of \ext\ from the emission of the nearby pulsar, we first performed a timing analysis on the \psr\ data to gate the pulsed emission, by means of the pulsar analysis package \textsc{TEMPO2} \citep{2006MNRAS.369..655H}. To obtain the phase curve of \psr, we assigned the corresponding phases to the gamma-ray events localized in a region of 0.6\degr{} of radius around the pulsar position (RA = 155.76\degr{}, DEC = -57.77\degr{}). Then, we computed the phases of the events using the updated ephemeris of \psr\ at epoch 55635 MJD. The new ephemeris that allows us to enlarge the data set to more than 11 years was provided by the \emph{Fermi}-LAT collaboration. Finally, we applied the Bayesian Blocks method \citep{2013Scargle} to the \psr\ light curve, and obtained different components, defined as ON, OFF, and {\it Bridge} emission. We computed the contribution to the phase curve of each of the sources in the model with the \emph{Fermi} tool {\it gtsrcprob}, which assigns to every event the probability of belonging to each source of the model.

The spectral energy distributions of the sources were obtained for 12 energy bins (spanning from 200\,MeV to 500\,GeV) using the whole ROI, with the \ext\ and \psr\ spectra characterized by power-law and exponential cutoff power-law models respectively. The sources SEDs are computed for the different components of the phase curve, cutting the events in phase. We tested the spectral analysis consistency studying the systematic uncertainties on the SEDs, mainly due to the LAT effective area ($\rm{A}_{eff}$) and the Galactic diffuse emission model. The systematic uncertainties due to the LAT effective area are computed with the bracketing $\rm{A}_{eff}$ method\footnote{\url{https://fermi.gsfc.nasa.gov/ssc/data/analysis/scitools/Aeff_Systematics.html}}, and the ones due to the diffuse Galactic model were tested by artificially changing the normalization of the same by $\pm 6$ per cent with respect to the best-fitted value \citep{Ajello_2011,2018Jian}.

On the other hand, the size of \ext\ was analyzed with the \textsc{fermipy} extension method, based on a likelihood ratio test with respect to the point-source hypothesis. We tested both 2D symmetric Gaussian and 2D radial disk models for the morphology of \ext. The best-fit extension in each case is computed performing a likelihood profile scan over the source width (68 per cent containment) and maximizing the model likelihood. 

Finally, we imposed different additional cuts of energy on the data described above (at 700\,MeV, 3\,GeV, 10\,GeV, 40\,GeV, 70\,GeV and 135\,GeV), and studied the morphological characteristics of \ext\ re-analysing the data in differential energy bins (see Figure \ref{fig:sizevsenergy}). 


\begin{figure}
  \centering
  \includegraphics[width=0.5\textwidth]{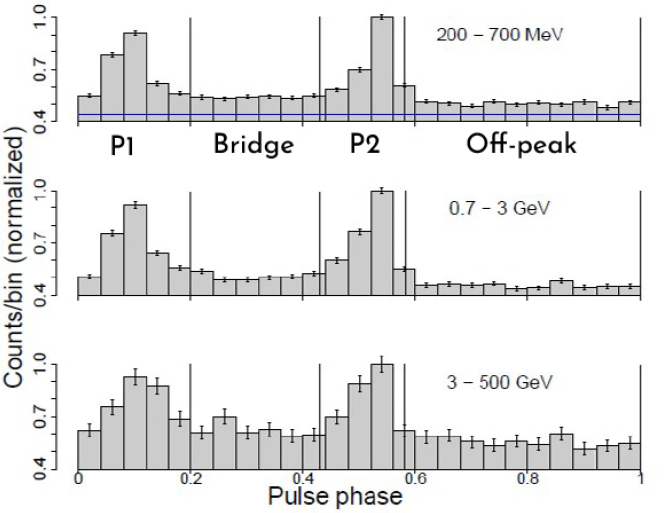}
  \caption{Phase curve of PSR J1023.1-5745 obtained at different energies in a region of 0.6\degr{} around the pulsar position (normalized by the height of P2) with the different components of the phase curve noted.
  The horizontal blue line corresponds to the expected contribution in the phase curve of background sources (i.e., all the sources in the field of view except for \ext\ and \psr).}
  \label{fig:phasecurve1023}
\end{figure}

\begin{figure*}
  \centering
  \includegraphics[width=0.45\textwidth]{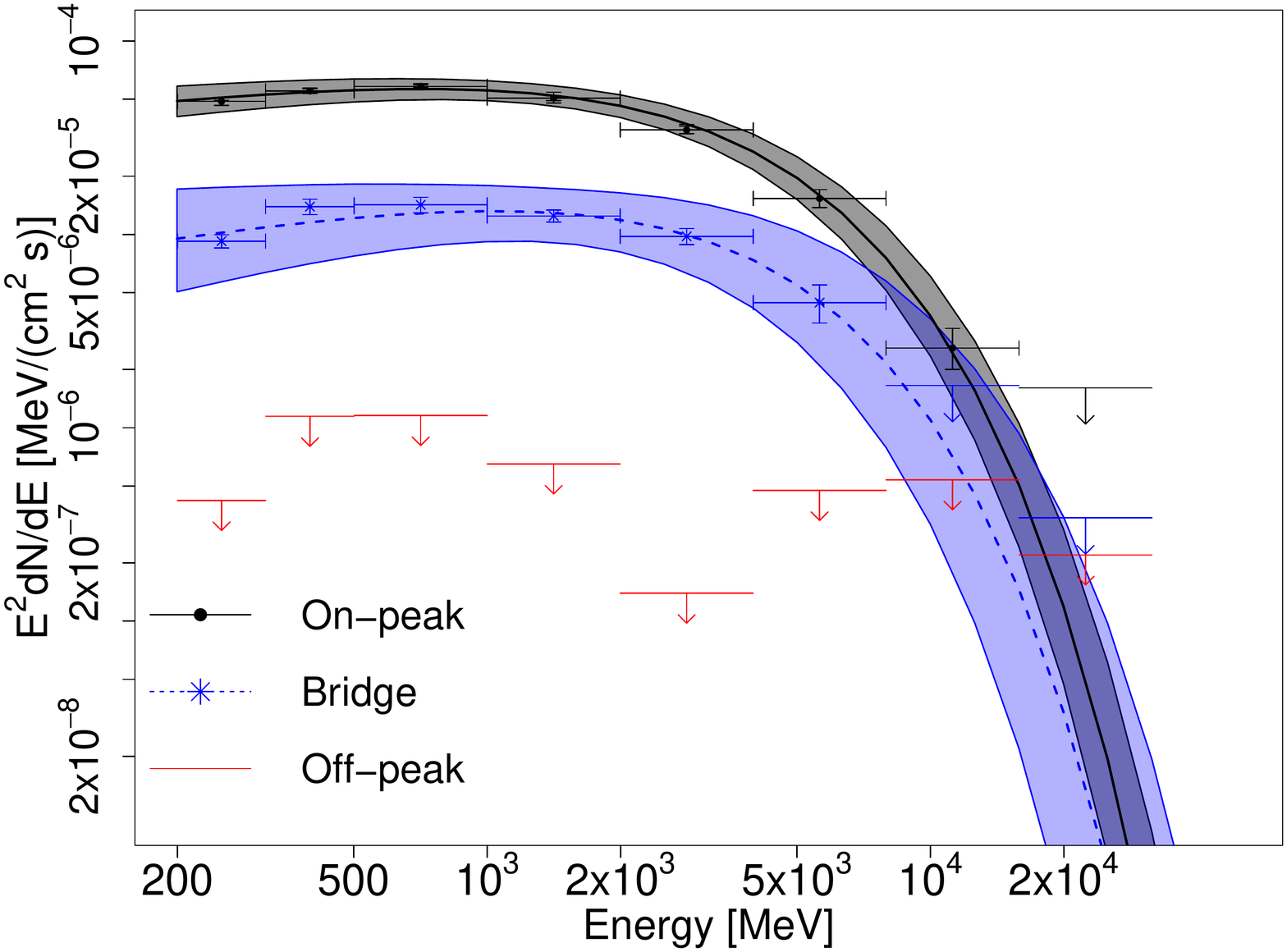}
  \includegraphics[width=0.45\textwidth]{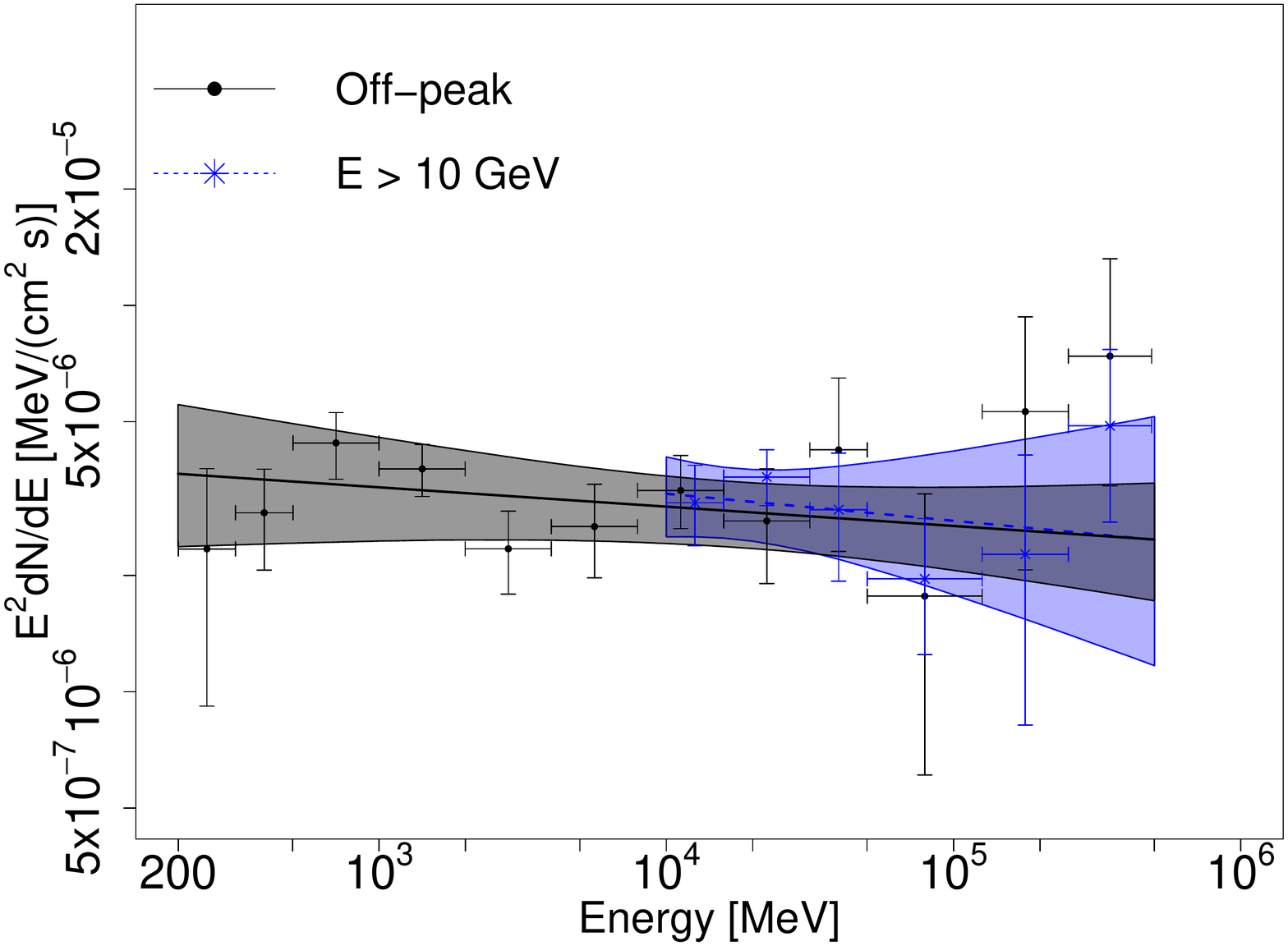}
  \caption{On the left a) Spectral energy distribution (with only statistical errors) of \psr\ derived from the on-peak (in black, with dot markers), the {\it Bridge} (in blue, with star markers), and off-peak-region (red upper limits). On the right b) Spectral energy distribution (with only statistical errors) derived from \ext\ after gating the pulsar emission (in black) and analysing only the gamma-ray emission above 10~GeV (without cutting in phase, in blue). The shaded area marks the 1$\sigma$ error on the fitted spectral model.}
  \label{fig:spectrums1023}
\end{figure*}

\section{Results}
\label{sec3}

The analysis of the pulsed emission from \psr\ with the Bayesian Blocks method showed two peaks, spanning from 0 to 0.2, and from 0.43 to 0.58 in phase, with a {\it Bridge} emission in between.  The peaks are centered at $\approx 0.09$ and $\approx 0.54$ in phase respectively and present a similar width ($\sigma \approx 0.02$ in phase if fitted with a Gaussian profile). The pulsed emission was analyzed in three intervals of energy: from 200\,MeV to 700\,MeV, from 700\,MeV to 3\,GeV, and above 3\,GeV. No significant shift was observed between the peaks at different energies. Also, the relation between the height of the peaks, and between the number of events in each peak ($\rm{P1}/\rm{P2} \approx 0.8$), does not differ significantly for the different energy bins. The second peak is the dominant one in the three intervals of energy. The number of events in the {\it Bridge} component (for the energy bins mentioned) is $\sim 10$ per cent larger than the one expected from the off-peak statistics. The three pulse profiles in the different energy bins are shown in Fig. \ref{fig:phasecurve1023} (counts per bin, for each phase). The Galactic diffuse emission is dominant both in the {\it Bridge} and off-peak regions. Hence, a similar level (in counts per bin) is observed in Fig. \ref{fig:phasecurve1023} for both components. The same contribution from \ext\ is expected in each bin of the phase curve (for a given energy interval), since the extended source emission should not vary in synchrony with the pulsed one.

\psr\ (described as a point-like source with an exponential cutoff power-law spectrum) is located at $\rm{RA} = 155.772\degr{} \pm 0.005\degr{}$ and $\rm{DEC} = -57.764\degr{} \pm 0.005\degr{}$ (see Table \ref{tab:position.}). The source is well-detected in the on-peak and {\it Bridge} regions with $\sqrt{TS} \gtrsim 80$ in both intervals. The results of the spectral analysis of \psr, performed both in the on-peak and {\it Bridge} intervals are summarized in Table \ref{tab:spectra.} (see Figure \ref{fig:spectrums1023} left panel). Note that the spectral energy distribution is characterized by an exponential cutoff located at an energy of $\sim 3$ GeV. Also, the cutoff energy obtained in the on-peak region is in agreement with the best fitted one for the {\it Bridge} component, while the spectral indices in both analyses are compatible within the uncertainties (at 95\% CL). In addition, the positions fitted for \psr, both in the on-peak and {\it Bridge} regions, are in agreement at 95\% CL. The point-like source, however, is not detected if selecting only off-peak events. In this region, only upper limits for the flux of \psr\ can be derived (see the red arrows in Figure \ref{fig:spectrums1023} left panel). This argues in favor of a reduced contribution of the pulsar emission in the defined 'OFF' region in comparison to the 'Bridge'. Admittedly, however, a close flux level between both such regions complicates this distinction.

To investigate the emission associated with the extended source \ext, we analyze the off-peak interval (from 0.58 to 1 in phase) to minimize the contamination from \psr. 
The analysis results in a significant detection of \ext\ ($\sqrt{TS} \approx 14$), with the source located at $\rm{RA}= 155.93\degr{} \pm 0.03\degr{}$ and $\rm{DEC} = -57.79\degr{} \pm 0.03\degr{}$ (see Table \ref{tab:position.}).
The parameters of the best-fitting power-law spectrum for \ext\ are summarized in Table \ref{tab:spectra.} (see Figure \ref{fig:spectrums1023} right panel).
The source spatial component is best described by a symmetric 2D Gaussian model, and the fit of the extension resulted in a 68 per cent containment radius of $r_{68} = 0.24\degr{}\pm 0.03\degr{}$ (i.e., an intrinsic size of $\sigma = 0.16 \pm 0.02$, see \citealt{2012Lande}), with a 0.29\degr{} upper limit for the 95 per cent containment radius. We also tested a symmetric 2D disk model, which, however, does not improve the log-likelihood with respect to the best-fit Gaussian model, obtaining a 68 per cent containment radius compatible with the extension above (i.e., $0.25\degr{} \pm 0.02\degr{}$). Also, the log-likelihood of the best-fit Gaussian model with an extension fixed to the value measured by H.E.S.S. ($\sigma = 0.18\degr{}$, \citealt{2011A&A...525A..46H}) is not significantly smaller than the one corresponding to the best-fit extension described above ($\Delta TS \approx 1.6$, as expected, given the similar extension of the GeV and TeV excesses). The extension and position of \ext\ were also computed for the on-peak and {\it Bridge} components, where an extended source model was fitted simultaneously to the \psr\ emission. No significant difference with respect to the results in the off-peak region was observed. The 'OFF' region of \psr\ has been analyzed with LAT data in previous works. In the second catalogue of LAT gamma-ray pulsars (2PC, see \citealt{2013ApJS..208...17A}), an unidentified source was reported in the off-pulse region, defined from 0.76 to 0.02 in phase. The analysis of this source showed some indication of spatial extension ($TS_{\rm ext} = 30$). Interestingly, the integrated flux for such source, computed from 100 MeV to 316 GeV of energy, i.e., $1.79 \times 10^{-8}$ cm$^{-2}$s$^{-1}$ (with large uncertainty) is compatible with the one derived (in this work) for \ext\ in the same regime (of energy), i.e., $(1.97\pm 0,49) \times 10^{-8}$ cm$^{-2}$s$^{-1}$ (with only statistical errors). Similarly, the off-peak region was analyzed in \citealt{2011ApJ...726...35A}, with 16 months of LAT data. In this case, the off-peak emission was detected only at energies above 10 GeV. Despite the off-pulse region (from 0.85 to 1.13 in phase, in the cited paper) differs significantly from the definition in this work, and the small amount of data used compared to the reanalysis we present, the integral flux of the source reported (from 10 GeV to 100 GeV of energy), i.e., $(4.6 \pm 2.2) \times 10^{-10}$ cm$^{-2}$s$^{-1}$ is also in well agreement with the flux derived for \ext\ in the same interval ($2.62\pm 0.37) \times 10^{-10}$ cm$^{-2}$s$^{-1}$. A more recent reanalysis of the detected off-pulse emission with 45 months of LAT data (see \citealt{Acero:2013xta}) reported a similar integrated flux above 10 GeV of energy.

The position of the sources, fitted for the different components of the phase curve, implies a separation between \psr\ and the centroid of \ext\ of $5.3\pm 1.1$ arcmin, with the pulsar position well within the extended source (given the size measured for the same, see Fig. \ref{fig:MWL}). In addition, the best-fitted position of \ext\ (in the off-peak region) is in good agreement with the one estimated by H.E.S.S for \hess\ \citep[i.e., RA $=155.85\degr$ and DEC $=-57.79\degr$ ,][]{2011A&A...525A..46H}, with a separation between the centroids of the sources of $3.3 \pm 1.4$ arcmin (account statistical errors only).

To further investigate the spectrum and morphology of the high energy part of \ext\, we analyzed the data set above 10\,GeV (see Table \ref{tab:spectra.} and Figure \ref{fig:spectrums1023} right panel), where additional cuts in phase are not necessary due to the exponential decrease of the pulsed emission above $\sim 3$\,GeV (of energy). We included the emission of \psr\ (modeled as a point-like source with the spectrum shown in Table \ref{tab:spectra.}) in this analysis, detected with low significance at energies above 10\,GeV ($\sqrt{TS} \approx 3$). The extension fitted in this case for \ext\ was $r_{68} = 0.23\degr{}\ + 0.03\degr{}\ - 0.02\degr{}$ (for a 2D Gaussian model). Note that both the position of the extended source and the best-fitting power-law spectral model are in good agreement with the ones obtained in the off-peak region (at 95\% CL, see Figure \ref{fig:spectrums1023} right panel and Table \ref{tab:position.}). Also, the spectral index, obtained both in the off-peak region and at energies above 10\,GeV, is compatible with $\Gamma = 2$, within the uncertainties (at 95\% CL).

To conclude, we studied the morphology of \ext\ (in the off-peak component) in six energy bins (with breakpoints in 200\,MeV, 700\, MeV, 3\,GeV, 10\,GeV, 40\,GeV, 70\,GeV, and 500\,GeV). For this purpose, we fitted the extension of \ext\ (measured for a Gaussian profile, i.e 68 per cent containment radius) as a function of the energy (see Figure \ref{fig:sizevsenergy}). Then, we tested an energy-dependent shrinking model of the form $\sigma \propto E^{-\alpha}$, but no significant variation of \ext\ extension was observed. 
The flux fitted for \ext\ in each bin of energy is compatible with the one expected from the spectrum fitted in the full range of energy (from 200\,MeV to 500\,GeV), and the spectral index of the best-fitting model is compatible with $\Gamma = 2$ (at $3\sigma$), for each bin of energy. However, the angular resolution of \emph{Fermi}-LAT ($\sim 0.8\degr{}$ at 1\,GeV, see \citealt{Abdollahi_2020}) limits the fit of the \ext\ extension for energies below few GeV's. Note that, since the \emph{Fermi}-LAT PSF is $\sim 0.2\degr{}$ at 3\,GeV (68 per cent containment radius), an imperfect diffuse modeling can bias the extension measured at lower energies. The large scale diffuse emission, dominant at low energies ($\sim 500$ MeV), limits the extension measurements performed in this regime, as seen in Figure \ref{fig:sizevsenergy} (note the upper limit in the lowest energy bin). The contamination from the same, however, is decreasingly relevant at higher energies.


\begin{table*}
\centering
\footnotesize
\caption{Best-fitting models for \psr\ and \ext. The units of $N_{0}$ are $\rm{cm}^{-2} \rm{s}^{-1} \rm{MeV}^{-1}$ for \psr\ and $\rm{cm}^{-2} \rm{s}^{-1} \rm{MeV}^{-1} \rm{sr}^{-1}$ in the case of \ext.}
\label{tab:spectra.}
\begin{tabular}{clclclclcl}
\hline
Parameter & \psr\ (On-peak) &  \psr\ ({\it Bridge}) & \ext\ (Off-peak)  \\ 
\hline
$N_{0}$ & $(2.51 \pm 0.18_{stat} \pm 0.24_{sys}) \times 10^{-11}$ & $(6.57 \pm 1.22_{stat} \pm 2.44_{sys}) \times 10^{-12}$ & $(1.02 \pm 0.14_{stat} \pm 0.16_{sys}) \times 10^{-14}$ \\ 
$\Gamma$ & $1.74 \pm 0.05_{stat} \pm 0.16_{sys}$ & $1.61 \pm 0.17_{stat} \pm 0.56_{sys}$ &$2.05 \pm 0.06_{stat} \pm 0.33_{sys}$ \\ 
$E_{0}$ [GeV] & $1.95$ & $1.95$  & $17$  \\ 
$E_{\rm{cutoff}}$ [GeV] & $2.74 \pm 0.24_{stat} \pm 0.39_{sys}$ & $2.7 \pm  0.57_{stat} \pm 1.2_{sys}$ & -\\ 
$\sigma$ [deg] & - & - & $0.16\pm 0.02$ \\ 
\hline
\hline
\end{tabular}
\begin{tabular}{clclcl}
\hline
Parameter &  \ext\ ($E > 10$\,GeV) & \ext\ (On-peak) & \ext\ ({\it Bridge}) \\
\hline
$N_{0}$ & $(1.09 \pm 0.16_{stat} \pm 0.04_{sys}) \times 10^{-14}$ & $(1.03 \pm 0.21_{stat} \pm 0.10_{sys}) \times 10^{-14}$ & $(9.83 \pm 2.0_{stat} \pm 0.5_{sys}) \times 10^{-15}$ \\ 
$\Gamma$ & $2.07  \pm 0.15_{stat} \pm 0.02_{sys}$ & $(2.01 \pm 0.16_{stat} \pm 0.26_{sys})$ & $(2.11 \pm 0.14_{stat} \pm 0.22_{sys})$ \\
$E_{0}$ [GeV] & $17$ & $19.6$ & $19.6$ \\
$E_{\rm{cutoff}}$ [GeV] &  - & - & - \\
$\sigma$ [deg] & $0.15\pm 0.02$ & $0.16 \pm 0.04$ & $0.16 \pm 0.03$  \\
\hline
\hline
\end{tabular}
\end{table*}












\section{Discussion}
\label{sec4}


The gamma-ray emission from the bright GeV pulsar \psr\ dominates the emission below a few GeV. Within the current statistics, the light curve remains similar in different energy ranges, contrary to other pulsars such as Crab or Vela. It is well described by two narrow peaks and a {\it Bridge} region between them. The spectrum obtained from this region is very similar to the one from the peaks, with an exponential cut-off at $\sim$3~GeV, indicating a common location of the radiation zone, most likely within the pulsar magnetosphere or its vicinity.   


The analysis of the gamma-ray emission, gating off the On- and {\it Bridge}-regions, combined with the large dataset, allow the detailed investigation of the spectral shape and morphology of the extended underlying source \ext. The measured size of the GeV source is in good agreement with the one measured by H.E.S.S. for \hess\ (see Fig. \ref{fig:MWL}). The spectrum obtained from the $\sim$0.2\degr region is hard ($\Gamma$=2) and connects smoothly the emission observed with LAT with the one in the TeV regime (see Fig. \ref{fig:piondecay}), indicating a clear identification of \ext\ as the GeV counterpart of \hess. The spectral results agree with the ones reported in \citealt{2017Ackermann}, but are in tension with the ones obtained by \citealt{2018A&A...611A..77Y}. For the latter, the ephemeris used was valid only for a reduced period in comparison to the temporal span of the data analyzed, resulting most likely in a contamination from the pulsar that affected the spectrum of the extended source. For this work, in turn, we have used the most updated ephemeris for the pulsar, valid for the entire data set analyzed.

The new characterization of the emission, thanks to the larger data set and better source discrimination, provides new clues to establish the origin of the gamma-ray emission. Different scenarios are discussed in the following, in the context of the new morphology and spectral features found.

\subsection{PWN scenario}
\label{sec4.1}

\begin{figure}
  \centering
  \includegraphics[width=0.5\textwidth]{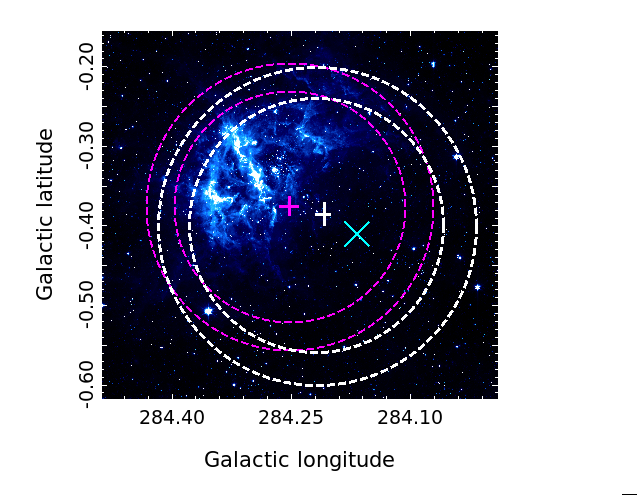}
  \caption{{\it Spitzer}/IRAC GLIMPSE Mosaic obtained from the GLIMPSE
  survey archival data. The magenta dashed lines correspond to the 1$\sigma$ interval for the best-fit extension of \ext, with the central cross marking the best-fit position (with the 1$\sigma$ error). The white dashed lines and cross correspond, similarly, to the morphological characteristics of the H.E.S.S. source \hess, as described in \citealt{2011A&A...525A..46H}. The cyan cross corresponds to the position of \psr. 
  }
  \label{fig:MWL}
\end{figure}

The discovery of the energetic pulsar \psr\ (with spin down energy $\dot{\rm E}=10^{37}$ erg s$^{-1}$), together with the proven efficiency of PWNe to produce TeV gamma-rays \citep{2018A&A...612A...2H} triggered the interpretation of \hess\ as a PWN energised by \psr. 
The association of the stellar cluster with the birth site of \psr\ ($\tau_{c}$=4.6 ky, \citealt{2010ApJ...725..571S}) is not straight-forward, implying an unrealistic transverse velocity of the pulsar \citep{2011ApJ...726...35A} if the same is located at the distance the cluster is believed to be $\sim$6 kpc \citep{2007ApJ...665L.163D}. Several time-dependent modelings of a Pulsar/PWN scenario have been proposed to connect \hess\ with \psr\ \citep{2011ApJ...726...35A,2018A&A...612A...2H}. In the reanalysis presented here, we found a noticeable morphological (see Fig. \ref{fig:MWL}) and spectral (see Fig. \ref{fig:piondecay}) overlap between the extended GeV source \ext\ and the TeV source \hess, which provides strong constraints on the PWN interpretation. In the same, the gamma-ray spectrum is expected to show a peak at energies just below the cut-off energy in the electron spectrum, in the TeV regime. The GeV gamma-ray spectrum due to the interactions of electrons with magnetic and radiation fields is effectively uncooled, up to the cut-off energy, since IC and synchrotron loss times are much longer in a typical PWN environment. The IC cooling time corresponding to the photon fields in the region where the pulsar is located (dust IR photons, stellar photons and CMB, see Table \ref{tab:pwn} where the density of the dust/stellar, $\omega_{\rm FIR/*}$ at temperature T$_{\rm FIR/*}$ is given) is $> 10 ^4$ yr, whereas a magnetic field of $> 400\ \mu$G should be considered to cool efficiently a few GeV electrons ($\tau_{\rm syn, yr}\simeq1.3\times10^7 B_{\mu \rm{G}}^{-2}E_{\rm{e,TeV}}^{-1}$). The uncooled $\Gamma_{\gamma} \sim 2$ spectrum extends from $\sim 200$ MeV to a few hundred GeV, where it meets the one measured by H.E.S.S. (see Figure \ref{fig:piondecay}). This photon spectrum would correspond to an electron spectrum of index $\alpha \approx2\Gamma_{\gamma}-1 =3$ (in the Thompson regime, \citealt{1970RvMP...42..237B}). Such a spectral index is difficult to reconcile with acceleration theories and typical injection spectra in PWNe. To further investigate this, we model the data using a PWN radiative scenario, in which the particle evolution and radiation are evolving in time according to the approach taken in \citealt{2014JHEAp...1...31T}. The time-dependent gamma-ray spectrum is obtained using the \textsc{GAMERA} software \citep{2015ICRC...34..917H}. Likewise, we used a broken power-law injection spectrum with similar shape to those found for young PWNe (see Table \ref{tab:pwn} for details and Table 2 in \citealt{2014JHEAp...1...31T}).
The time-dependent modeling used does not account for morphological changes (i.e., assumes energy-independent morphology).
The best model representing the data is shown in Fig. \ref{fig:piondecay}, integrating in time up to the estimated pulsar age ($\tau_c\sim $4.6 kyr). The upper limit on the X-ray emission was obtained by \citealt{2009PASJ...61.1229F} using {\it Suzaku} observations (in a $17\arcmin.8 \times 17\arcmin.8$ field). The disagreement between the LAT data and the model is evident, especially at the lower energies, and one should then consider a more complex modeling, with more than one electron population, to obtain a good representation of the data. According to \citealt{2007A&A...474..689M}, the stellar average density associated with Westerlund 2 could be as high as $500$ eV/cm$^{3}$ at a  temperature of $3 \times 10^{4}$ K, well beyond the nominal $2$ eV/cm$^{3}$ density used in the model described in Table \ref{tab:pwn}. However, we found no energy density in the cited range for which a better fit than that depicted in Figure \ref{fig:piondecay} is derived (see Figure \ref{fig:refereeplot}). A second possibility to flatten the spectrum involves the contribution of different stellar photon fields, which would be indeed expected from a region like Westerlund 2 \citep{2013AJ....145..125V,1995fmw..conf..183G}. However, the morphology of such a source, peaking at the regions of high stellar radiation density, would differ from the one observed, rendering this possibility unlikely. Additionally, the similar extension between the LAT and the H.E.S.S. measurements disfavors further the PWN scenario, where usually a larger GeV nebula, due to cooling or/and energy-dependent diffusive transport, is observed \citep{2020A&A...640A..76P,2019A&A...621A.116H,2012A&A...548A..46H}.  

\begin{table}
\centering
\scriptsize
  \caption{Physical parameters of \psr\ and its putative PWN. The pulsar rotation parameters $f$ and $\dot{f}$ are obtained from ATNF catalogue. The braking index $n$ and ejecta mass $M_{ej}$ are fixed, following the results from \citealt{2014JHEAp...1...31T}. The broken power-law spectral shape of electrons (defined with the indices $\alpha_{1,2}$ and breaking energy $\gamma_b$), and magnetic field in the region ($B$), that best represent the data, are listed in the third section of the table.}


  \begin{tabular}{  l l}
 
 \hline
{ {Pulsar \& Ejecta}}  \\
 \hline
 
$f$ (Hz) & 8.97 \\

$\dot{f} (-10^{-12}\ $Hz s$^{-1}$)& 30.88 \\

$\tau_{c}$   (kyr)  & 4.6 \\

$L(t_{age})$  (erg/s) 	& 1.1 $\times 10^{37}$ \\

$n$ &  2.509 \\

$D$  (kpc)  & 5 \\   

$M_{ej}$ ($M_{\odot}$)                    	&  10 \\

\hline
{ Environment }\\
\hline

$T_{FIR}$ (K)                                  	& 30 \\

$w_{FIR}$ (eV/cm$^{3}$)       	&   1 \\

$T_{*}$ (K)     	&   2500\\

$w_{*}$   (eV/cm$^{3}$)	& 2 \\

$n_{H}$    	&   1.0         \\

 \hline
{ Particles and field } \\
 \hline

$\gamma_{b}$                      & $5\times 10^5$ \\

$\alpha_{1}$              & $1.5$ \\

$\alpha_{2}$              & $3.0$ \\

$B(t_{age})$    ($\mu$G)          & $7$ \\

\hline
\end{tabular}
\label{tab:pwn}
\end{table}

\subsection{Scenarios related to Westerlund 2 massive stars}
\label{sec4.2}

Electrons can also be efficiently accelerated in open clusters via shocks in e.g. massive stars \citep{2014PhRvD..90j3008B}, generating gamma-rays via IC or Bremsstrahlung radiation. However it is expected that the resulting gamma-rays should correlate with the region of high photon density. The gamma-ray emission extends up to $\sim25_{d_{5 kpc}} $pc, in contrast with the $\sim4_{d_{5 kpc}} $pc core radius of the cluster, disfavoring a leptonic origin of the GeV and TeV emission observed. Likewise, the source size remains constant, within the errors, for different energy bands, which indicates a stable dependency of the cooling time with energy, contrary to what is expected in a leptonic scenario.
 
\begin{figure*}
  \centering
  \includegraphics[width=\textwidth]{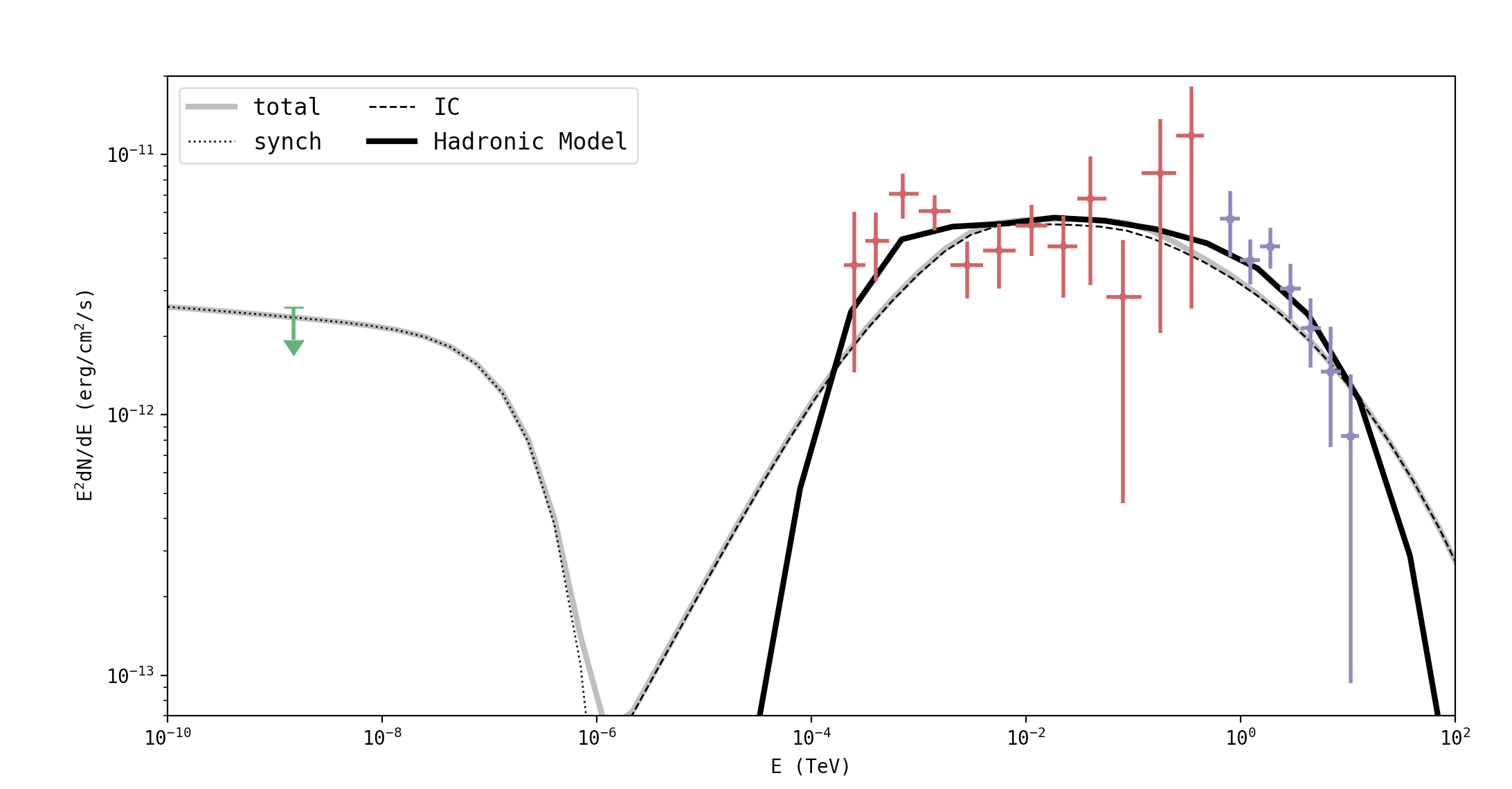}
  \caption{Best-fitting models for \ext\ spectrum in the pion decay (solid line) and PWN (dashed line) hypotheses. The red points correspond to the \emph{Fermi}-LAT data analysis (with only statistical errors), and the purple ones to H.E.S.S.  \citep{2011A&A...525A..46H}. The upper limit in the X-ray domain is obtained from \citealt{2009PASJ...61.1229F}.}
  \label{fig:piondecay}
\end{figure*}

\begin{figure}
  \centering
  \includegraphics[width=0.55\textwidth]{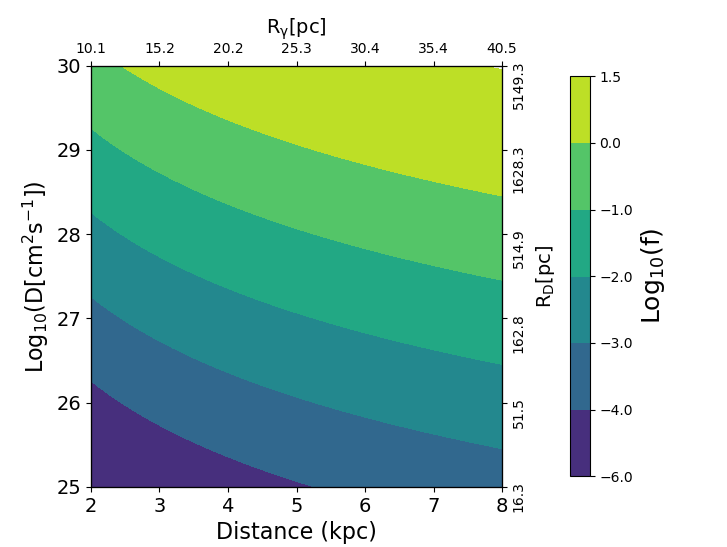}
  \caption{Acceleration efficiency ($f$) obtained with respect to the distance and diffusion coefficient ($D$) assumed (for a fiducial target mass in molecular clouds of $4.5\times 10^{5}\ M_{\odot}$). The upper and right axes correspond to the estimated gamma-ray source and CRs diffusion radius, respectively.}
  \label{fig:efficiencyplot}
\end{figure}

In contrast, an hadronic interpretation fits naturally the hard spectrum found from  $\sim$200 MeV up to a few tens of GeV, where it connects smoothly with the H.E.S.S. spectrum, and continues up to a few TeVs before showing a drop of the flux. Next, we constrain the proton population that powers the gamma-ray source by modeling the 200 MeV to 20 TeV emission using the {\sc naima} package (version 0.8.4, \citealt{naima}). 
To calculate the SED, we used a distance of 5 kpc. The molecular content in the region has been deeply investigated by several authors \citep{2007ApJ...665L.163D,2014ApJ...781...70F} using millimeter wave CO spectroscopy. Several massive molecular clouds were found within the surrounding of Westerlund 2. The total mass is estimated to be between $(1.7-7.5)\times10^{5}M_{\odot}$. We used a particle distribution described by a particle index $s$ and an amplitude $N_{p}$, up to an energy cut-off $\rm{E}_{cutoff}$. The corresponding gamma-ray emission due to pion decay radiation is calculated using the parametrization in \citealt{2014PhRvD..90l3014K} implemented in \textsc{naima}, and compared to the experimental data. 
The best-fitting model for the joint \emph{Fermi}-LAT and H.E.S.S. data corresponds to an exponential cut-off power-law proton spectrum with cut-off energy $\rm{E}_{cutoff} = 93 \pm 8$ TeV, and a particle index of $s = 2.09 \pm 0.01$, referenced to 1\,TeV, 
see Fig. \ref{fig:piondecay}). The fit to the data is done by means of a Log-Likelihood approach. 
For a distance of 5 kpc, the total energy in protons estimated above 1.22\,GeV (the threshold kinetic energy for pion production in pp interactions) is $W_{\rm p} = (1.3 - 5.9)\times10^{48}$ erg, for densities of n = (7.5 - 1.7)$\times10^5$ M$_\odot$/$V_{24pc}$, as derived by \citealt{2014ApJ...781...70F} and \citealt{2007ApJ...665L.163D}, respectively. We estimated a lower limit for the cut-off energy on the proton spectrum of $\approx 37$ TeV (at 95\% CL), by comparing the maximum likelihoods of the data obtained for the exponential cutoff power-law and power-law models with the likelihood-ratio test.

The total energy in protons (above 1.22\,GeV) can be compared with the total mechanical power of the stellar winds in the Westerlund 2 cluster:  $\rm{W}_{\rm{tot}}$ =$f L_0T_0$, which results in a modest acceleration efficiency of $f=10^{-4}$ ($10^{-6}$ / $5\times 10^{-3}$), for the well known age ($T_0 =2\times10^6$ yr), a distance of 5\,kpc (2\,kpc / 8\,kpc) and the available energy budget in the form of kinetic energy of stellar winds ($L_0 = 2 \times 10^{38}$ erg s$^{-1}$). 

If we consider the total volume defined by the size of the GeV source ($\sim24$ pc, i.e., $V_{24\rm{pc}}$ of volume for a spherical source), the energy density of protons in the region is $\omega_{\rm{p}}$  = (1.3-5.9)$\times10^{48}\  \rm{erg}/V_{24\rm{pc}}\simeq (0.5-2.2) \ eV/cm^3$ (or 0.1-5.6 \ eV/cm$^3$ if considering the uncertainty due to the distance to the source), which is comparable to the density of protons found in others massive clusters \citep{2019NatAs...3..561A}. 
The maximum energy of this proton population is constrained by the best fit, described above, to be $\sim$90\,TeV, with a lower limit of 37\,TeV on the energy cut-off. 
It would be possible however that a significant part of these CRs has already escaped from the gamma-ray emission region, where the molecular content is enhanced, and therefore also the gamma-ray radiation. Under the assumption of a continuous injection of protons and spherical expansion \citep{1996A&A...309..917A}, the relation between the observed energy emitted by protons ($W_{em}$) and the total energy (in protons) available ($W_{tot}$) is: 
\begin{equation}
W_{em} / W_{tot} = (R_\gamma/R_D)^2
\end{equation}
where $R_\gamma$ and $R_D$ are the gamma-ray source and diffusion radii, respectively. Then, using a diffusion coefficient as in the ISM, i.e., $\rm{D}\sim 10^{28} \rm{cm}^2\rm{s^{-1}}$ and $R_{\rm D}=2\sqrt{T_0D}=515\,\rm{pc}$, we obtain that the total energy released in the form of CRs could reach $W_{\rm tot}\sim5\times 10^{50} \rm{erg}$ (for a distance of 5\,kpc), which is still a few percent of the total available energy in the kinetic winds ($L_{0}T_{0} \sim 10^{52}$ erg). Note that this number is affected by the uncertainties in the distance: for instance, the efficiency obtained is larger than 10\% for a distance of 8\,kpc with the quoted diffusion coefficient of the ISM. Likewise, formally, the value of the diffusion coefficient could be larger, and correspondingly, the CR halo could reach up to 5\,kpc (for $D \sim 10^{30} \rm{cm}^2\rm{s^{-1}}$). These uncertainties have an effect on the total efficiency, reaching in some extreme cases an unrealistically large fraction of the total energy to be transferred to CRs.

If the region is, instead, affected by large turbulence, expected in the surrounding of an accelerator, the CRs diffusion could be much slower \citep{2013ApJ...768...73M,2020Schroer}. If the size observed at GeV and TeV energies ($R_{\gamma} \sim 24\, (d/\rm{5kpc}$) pc) reflects the propagation depth of CRs, the diffusion coefficient would be much lower than in the ISM $D \sim 3\times 10^{25} \rm{cm}^2\rm{s^{-1}}$ (or (0.4   - 6)$\times$10$^{25}\rm{cm}^2\rm{s^{-1}}$ for 2 and 8 kpc respectively), which is in tension with the value of the diffusion coefficient at these energies in the Bohm regime  \citep{2019NatAs...3..561A}. That points to a certain CR halo around Westerlund 2, beyond the size traced by LAT and H.E.S.S., that could, in principle, extend up to a few tens of parsecs. The discussion above is roughly summarized in Fig. \ref{fig:efficiencyplot}. 
More precise estimations of the distance to the cluster, foreseen with {\it Gaia} DR3 \citep{2018AJ....156..211Z} (and therefore of the real gamma-ray size), would provide constraining limits on the diffusion of CRs around the source, for a range of acceleration efficiency $f$.

Another possibility to explain the gamma-ray emission involves energetic SNR explosions within the cluster, injecting CRs in the surrounding \cite{2019ApJS..244...28T}. The lack of shell-like structure and the high efficiency required in the case of diffusive CRs (50$\%E_{\rm SN}$, for a standard $E_{\rm{SN}}=10^{51}$erg) render this hypothesis less attractive, at least for a single SNR event, than the one that attributes the origin of CRs to the stellar winds.

\section{Conclusions}
\label{sec5}

The reanalysis of the large LAT dataset presented here results in a clear identification of the extended source \ext\ with the TeV source \hess. The matching spectral and morphological agreement, with no signs of cooling features in the size of the source, points to a common origin of the radiation. The combination of the two results obtained, that is, the extended source beyond the cluster size and in a good agreement with the TeV radiation, and the hard spectrum that continues towards low energies, constitutes evidence of the hadronic nature of the gamma-ray emission detected using \emph{Fermi}-LAT and H.E.S.S. data. 
Here we evaluate the different scenarios proposed to explain \hess, in light of the new information provided by the LAT spectrum and concluded that the gamma-ray source is compatible with being of hadronic origin, and related to the Westerlund 2 stellar cluster, rather than to leptonic emission from either the PWN associated with \psr, or the cluster itself. However, the PWN scenario, even if unlikely, can not be conclusively ruled out when considering the uncertainties in the data points. Moreover, the reasonable uncertainties of both models as well as the possible existence of a multi-component photon energy density introduce further ambiguity. 

The results presented, pointing to the Westerlund 2 cluster as a hadronic accelerator with a hard $\sim2$ spectral index, indirectly support the hypothesis of stellar clusters as significant contributors to the GCR sea. In particular, these CRs steaming from Westerlund 2 might also extend to a very large halo ($\sim$200\,pc radius) around the cluster, as proposed in \citealt{2018A&A...611A..77Y}. The total energy in protons we derive ($\sim5\times 10^{50} \rm{erg}$) can easily account for the total luminosity observed, requiring an acceleration efficiency in the cluster of $f=0.04$, which is still moderate for acceleration theories in wind shocks. The spectrum at TeV energies seems to change the hard 2 index trend found in the LAT data, softening towards higher energies. This spectral shape would imply a low energy cut-off, disfavoring Westerlund\,2 as a PeVatron accelerator. Deeper observations with H.E.S.S. or with sensitive TeV instruments in the South such as CTA \citep{2019scta.book.....C} in the future should provide a definitive answer to the PeVatron nature of \hess.

\section*{Data Availability}
The \emph{Fermi}-LAT data underlying this article are available at \url{https://fermi.gsfc.nasa.gov/ssc/data/access/lat/}. The GLIMPSE survey archival data underlying Figure \ref{fig:MWL} are available in \url{https://irsa.ipac.caltech.edu/data/SPITZER/GLIMPSE/}.

\section*{Acknowledgements}
This research was supported by the Alexander von Humboldt Foundation (EdOW). E. M. \& D. F. T. acknowledge the support of the grants AYA2017-92402- EXP, PGC2018-095512-B-I00 (Ministerio de Econom\'ia y Competitividad) and SGR2017-1383 (Generalitat de Catalunya). We acknowledge the support of the
PHAROS COST Action (CA16214). We made use of R Project for Statistical Computing \citep{Rmanual}. The \emph{Fermi}-LAT Collaboration acknowledges generous ongoing support from a number of agencies and institutes that have supported both the development and the operation of the LAT as well as scientific data analysis. These include the National Aeronautics and Space Administration and the Department of Energy in the United States, the Commissariat a` l’Energie Atomique and the Centre National de la Recherche Scientifique / Institut National de Physique Nucl\'eaire et de Physique des Particules in France, the Agenzia Spaziale Italiana and the Istituto Nazionale di Fisica Nucleare in Italy, the Ministry of Education, Culture, Sports, Science and Technology (MEXT), High Energy Accelerator Research Organization (KEK) and Japan Aerospace Exploration Agency (JAXA) in Japan, and the K. A. Wallenberg Foundation, the Swedish Research Council and the Swedish National Space Board in Sweden. This work
performed in part under DOE Contract DE-AC02-76SF00515.



\bibliographystyle{mnras}
\bibliography{wd2}




\appendix

\section{Analysis details}

\begin{figure}
  \centering
  \includegraphics[width=0.5\textwidth]{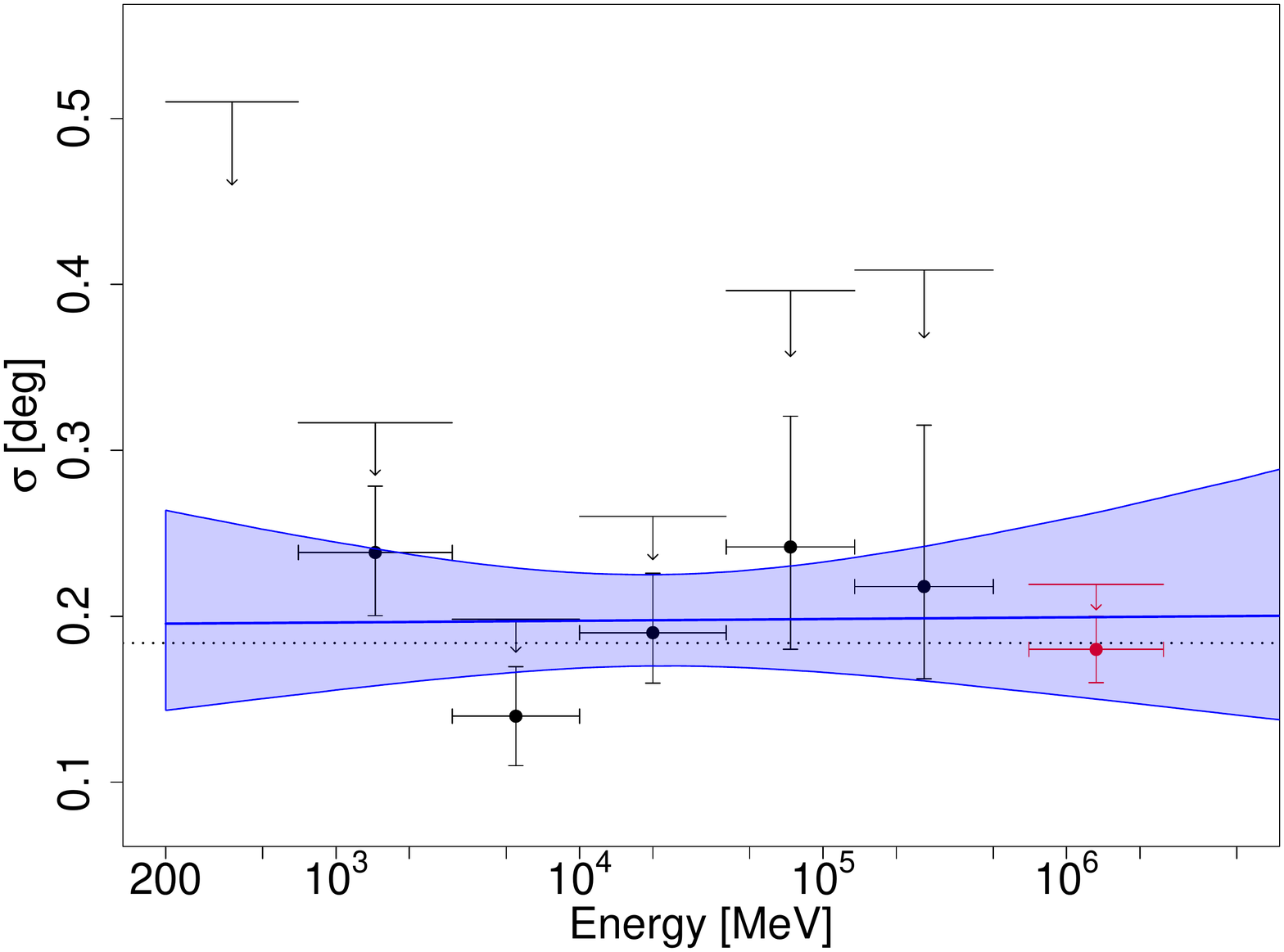}
  \caption{Extension of \ext\ (i.e., intrinsic width for a Gaussian fit) measured for different energy bins (in black points for LAT data, with only statistical errors). The red point corresponds to the intrinsic extension measured by H.E.S.S. from 0.7 TeV to 2.5 TeV of energy \citep{2011A&A...525A..46H}. The best-fitted power-law model (for the joint LAT and H.E.S.S. data) is plotted in blue line with the $1\sigma$ region noted (the blue shaded area), together with the weighted mean size (black dashed line). The arrows correspond to an upper limit for the extension (i.e., $\sigma$, at 95\% CL), except for the one in the first energy bin corresponding with the upper limit for the 95\% containment radius ($r_{95}$).}
  \label{fig:sizevsenergy}
\end{figure}



\begin{table*}
\centering
\footnotesize
\caption{Best-fitted position (in equatorial and Galactic coordinates) for \psr\ and \ext\ (in degrees, with only statistical errors).}
\label{tab:position.}
\begin{tabular}{clclclclcl}
\hline
Parameter & \psr\ (On-peak) &  \psr\ ({\it Bridge}) & \ext\ (Off-peak) & \ext\ ($E > 10$\,GeV) \\
\hline
RA & $155.772 \pm 0.005$ & $155.79 \pm 0.01$ & $155.93 \pm 0.03$ &  $155.93 \pm 0.02$\\ 
DEC & $-57.764 \pm 0.005$ & $-57.77 \pm 0.01$ & $-57.79 \pm 0.03$ & $-57.76 \pm 0.02$ \\
$l$ & $284.168 \pm 0.005$ & $284.18 \pm 0.01$ & $284.25 \pm 0.02$ & $284.24 \pm 0.02$\\
$b$ & $-0.401 \pm 0.005$ & $-0.40 \pm 0.01$ & $-0.38 \pm 0.03$ & $-0.35 \pm 0.02$\\
\hline
\hline
\end{tabular}
\begin{tabular}{clclcl}
\hline
Parameter & \ext\ (On-peak) & \ext\ ({\it Bridge}) \\
\hline
RA & $155.98 \pm 0.04$ & $155.83 \pm 0.03$ \\ 
DEC & $-57.77 \pm 0.03$ & $-57.81 \pm 0.03$ \\
$l$ & $284.27 \pm 0.05$ & $284.22 \pm 0.03$ \\
$b$ & $-0.35 \pm 0.03$ & $-0.43 \pm 0.03$ \\
\hline
\hline
\end{tabular}
\end{table*}


\begin{figure*}
    \centering
    \includegraphics[width=\linewidth]{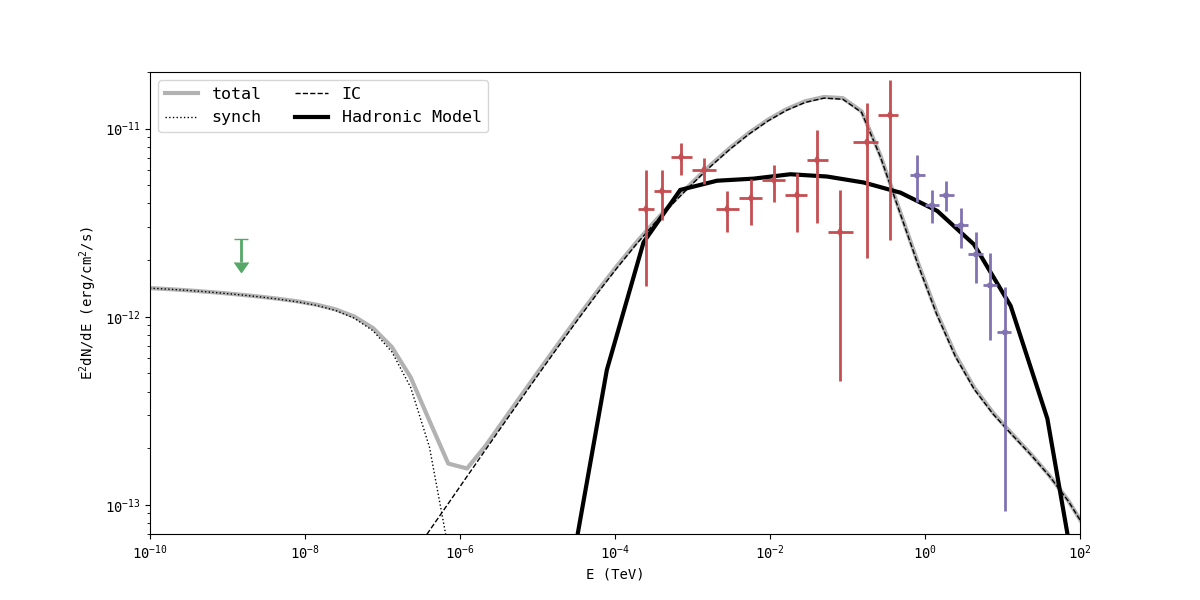}
    \caption{The same as Figure \ref{fig:piondecay}, but the best-fitting model for the PWN hypothesis depicted (dashed line) corresponds to a maximum stellar photon field of $w_{*} = 500$ eV/cm$^{3}$ at a temperature of T $= 3 \times 10^{4}$ K as in \citealt{2007A&A...474..689M}. }
    \label{fig:refereeplot}
\end{figure*}


\bsp	
\label{lastpage}
\end{document}